\documentclass[conference]{IEEEtran}

\usepackage{mathptmx}       
\usepackage{helvet}         
\usepackage{courier}        
\usepackage{type1cm}        
\usepackage{makeidx}         
\usepackage{graphicx}        
\usepackage{multicol}        
\usepackage[bottom]{footmisc}

\usepackage{amsmath,amssymb,amsfonts}
\usepackage{algorithmic}
\usepackage{graphicx}
\usepackage{textcomp}

\begin{document}

\title{Secure Indoor Location For Airport Environments\\
}

\author{\IEEEauthorblockN{Iv\'an Santos-Gonz\'alez, Alexandra Rivero-Garc\'ia and Pino Caballero-Gil}
\IEEEauthorblockA{\textit{Department of Computer Engineering and Systems} \\
\textit{University of La Laguna}\\
Tenerife, Spain \\
\{jsantosg, ariverog, pcaballe\}@ull.edu.es}
}

%
%
\maketitle

\begin{abstract}

This work presents a secure novel solution based on inertial measurement units to provide indoor location and positioning in airports. The use of different technologies allows to locate people with precision in this kind of indoor places where the use of the GPS is not possible. The system has been developed thinking in the low cost and in a possible future expansion of this kind of systems to improve the Quality of Service of the users in airports. The use of QR codes and low cost IMU devices through the use of people smartphones ensure this premise. An Android application has been developed to show the applicability and performance of the system. The security in this kind of systems is essential. This kind of systems needs to avoid the traceability of the IMU devices when users are using it. To solve this problem, the FourQ elliptic curve has been used to generate a shared key using the elliptic curve Diffie-Hellman protocol. The key generated with the FourQ is used then to cipher all communications through the use of the SNOW 3G stream cipher. The developed system offers promising results.
\end{abstract}


\section{Introduction}
\label{sec:intro}

Nowadays the use of positioning system is widespread. It is becoming more and more common that people do not know how to orient themselves with a simple map. It is well known that  outdoors the use of GPS positioning system provides an accurate location, but indoor the GPS system not offers good precision. The traditional way to solve this problem has been by putting static maps in different points of the building indicating where  you are inside the building map. The main disadvantage of this kind of information points is that they are not accessible in all building spaces.

When the time is an essential factor, as it is the case of the airports, the use of an accurate system is vital. To offer a good solution, people must use a device that they know and know how to use without prior training. It is known that smartphones are becoming more and more essential in our daily lives because we
do not use the smartphone only to make phone calls or to send  Short Message Service (SMS) messages, but also to do other tasks such as taking
pictures, recording videos, reading mails, locating with Global Positioning System (GPS) or surfing the Internet. Due to this, different indoor location solutions based on smartphones have been proposed in the last years.

The presented proposal is based on the use of two technologies: Quick Response (QR) codes \cite{QR}, which provides a short-range positioning, and an Inertial Measurement Unit (IMU) \cite{morrison1987inertial} settled on the user foot, which provides inertial changes to track the user's movement. The use of an Android application in combination with the measurements obtained by the IMU and the initial position obtained from the QR code, allows to provide real-time position in the application over the indoor map of the airport. Traditionally, the security of this kind of system has not been study. In this work, a new security scheme for this kind of system has been proposed, concretely to address the traceability problem. The use of the elliptic curve Diffie-Hellman protocol \cite{koblitz1987elliptic} through the use of the FourQ elliptic curve allows to generate a shared key between the smartphone and the IMU. Then, the Snow 3G stream cipher algorithm \cite{kircanski2011sliding} is used to encrypt all communications.

This  work is structured as follows. Section 2 describes some preliminaries. The proposed system is defined in Section 3. Section 4 introduces some features of the system security. Finally, some conclusions and open issues close this paper.

\section{Preliminaries}
\label{sec:preliminaries}

During the last years, different proposals have been presented  to solve the problem of indoor location in airports.  One of the most recent is the proposal in \cite{BoqueIndoorAeropuertos} where the authors propose the use Bluetooth beacons to, in combination with use of a smartphone with an Android application, locate users in the airport map. This is a good starting point, but in some cases, the precision of system based on the RSSI of any wireless technology suppose a huge problem. This is the case of the electromagnetic interferences, produce for example by a human body. It is well known that airports are very busy places, so the precision of this kind of system in airports could not be very precise. On the other hand, different smartphones have different wireless chipset, so two different smartphones have to measure different values of the beacon RSSI.
 
Traditionally, IMUs have been used to track the movement and/or position of users in different situations. There are different IMU types, but the ones used to track the movement and/or position usually have been the 6 Degrees of Freedom (DoF) or 9 DoF IMUs. A 6 DoF IMU usually has  a 3 DoF accelerometer and a 3 DoF gyroscope. The accelerometer is used to measure the acceleration on IMU movements in the x, y and z coordinate systems, that can be easily transformed into speed through the first time integral of the acceleration, and to position through the second time integral of the acceleration. Thus, it can be used to measure changes in the speed and position respectively. A problem that usually appears when obtaining speed and position through the use of the integral is that if the intrinsic constant error is not removed from the original measurement, the acceleration, it becomes a lineal error in the speed and in a quadratic error in the position, fact that would do the system unusable. The gyroscope measures the orientation in the x, y and z coordinate systems. A 9 DoF IMU has the 3 DoF accelerometer and gyroscope and adds a 3 DoF magnetometer, a sensor that measures the magnetic field and that  is usually used to get the global orientation due to the earth magnetic field. A complete guide of the most common error sources of the use of IMU for positioning systems and its effects on the navigation performance can be found in \cite{flenniken2005characterization}. 

A method that has been used usually based on  the measurements of these aforementioned sensors is the Dead-Reckoning. This method consists on the use of different algorithms based on easy trigonometric equations to get the actual position of an object or person, through operations based on the course and navigation speed. There are multiple algorithms that implement the Dead-Reckoning  \cite{Luna2015}. In that paper, a comparative study of different Pedestrian Dead-Reckoning algorithms is presented. A Pedestrian Dead-Reckoning algorithm is basically an algorithm that estimates the movement of a person by detecting steps, estimating stride lengths and the directions of motion. The results obtained in that work shows how this technique offers promising results with an average rate on the stride length estimation errors of about 1\% and an estimation below  5\% in the total travelled distance. Another method that is usually applied to improve the performance and to reduce the drift error on the sensors measurements is the use of static and adaptative filters, and one of the most used filters is the Kalman filter \cite{van2001square}. 

\section{Positioning System}
\label{sec:proposed}

The developed system consists on an Android application that shows in the indoor map of the airport the current position of the user. The system uses two different technologies to perform this feature. 

On the one hand, the QR codes are used at the entrance or some specific points of the airport to set the origin point of the user in the map. This technology has been used in the system because it is a short range communication technology with no error or small error in the initial position estimation. Another important aspect of this technology is that it is the cheaper technology in the world because we need only a paper with the code printed.

On the other hand, the use of an IMU located on the user foot supposes an static reference point. The use of an IMU located on the user foot is a more accurate way of collecting data than the smartphone because it is static and produces less noise than the use of  smartphone sensors. The IMU is used to collect data about the accelerometer, gyroscope and magnetometer sensors, which are sent to the user smartphone through the use of Bluetooth Low Energy  (BLE) \cite{gomez2012overview} technology. The use of the IMU unit instead of the user smartphone is due to the smartphone movements could add some noise to the measurements and the measurements obtained through the IMU sensors are more precise than the smartphone measurements. In the user smartphone, the sensor data are processed through the use of the Madgwick algorithm. This algorithm provides an accurate orientation of the user in a quaternion form \cite{horn1987closed}, which provides an absolute orientation from a relative one. Then, the quaternion is used to orientate the position in the indoor map.

The main part of the indoor positioning system is the part related to the IMU collected data and its treatment. In our positioning system a Metawear CPRO IMU that collect measures of the 3 sensors, accelerometer, gyroscope and magnetometer, obtaining $g$, $ degrees/seconds$ and $Tesla$ units respectively, has been used. The complete specifications of the IMU unit  can be shown in \cite{metawear}. This IMU unit transmit the data through BLE. The collected data is transmitted in real time to the smartphone where the treatment of the different variables is performed, including the conversion of accelerometer units, $g$,  to $m/s^2$, and the gyroscope units, $degrees/seconds$, to $rad/s$. Then, the Madgwick filter is applied to obtain the quaternion that represents the pitch yaw and roll. With these data, the step detection and the step length, the user's position is shown  over the map every time he/she takes a step.

\begin{figure}[!ht]
\begin{center}
\includegraphics[width=1\linewidth]{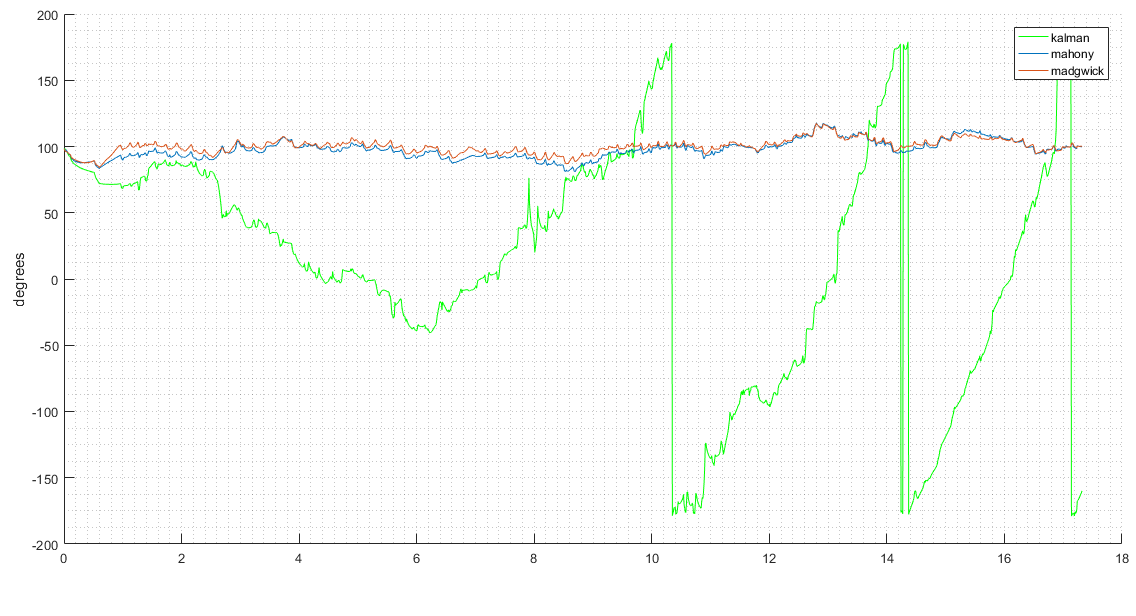}
\caption{Filters Yaw} 
\label{fig:yaw}
\end{center}
\end{figure}

\begin{figure}[!ht]
\begin{center}
\includegraphics[width=1\linewidth]{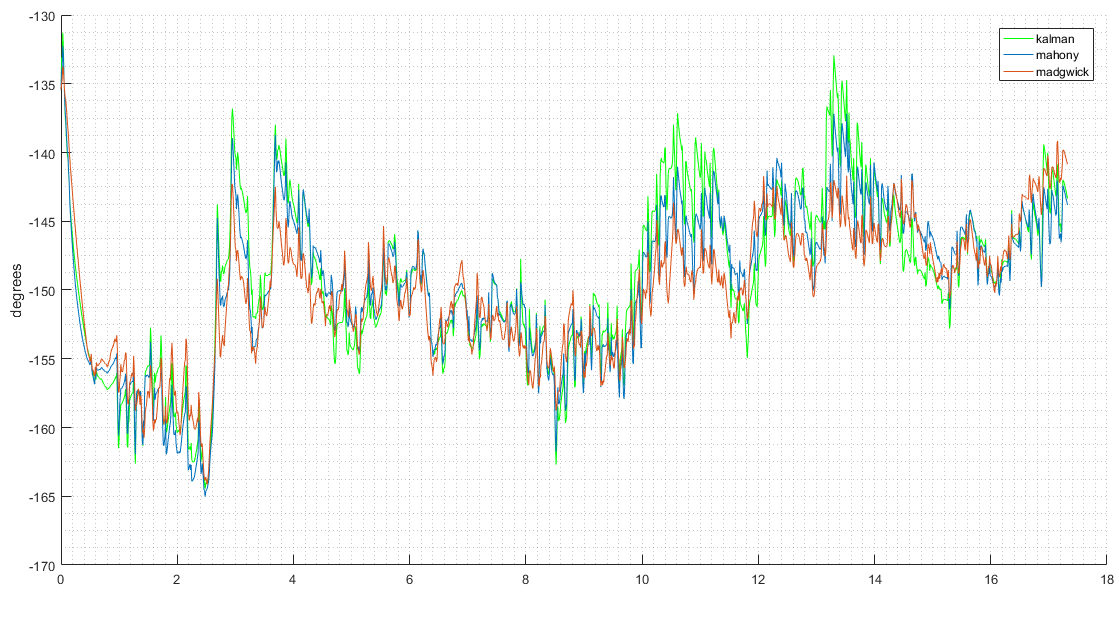}
\caption{Filters Pitch} 
\label{fig:pitch}
\end{center}
\end{figure}

\begin{figure}[!ht]
\begin{center}
\includegraphics[width=1\linewidth]{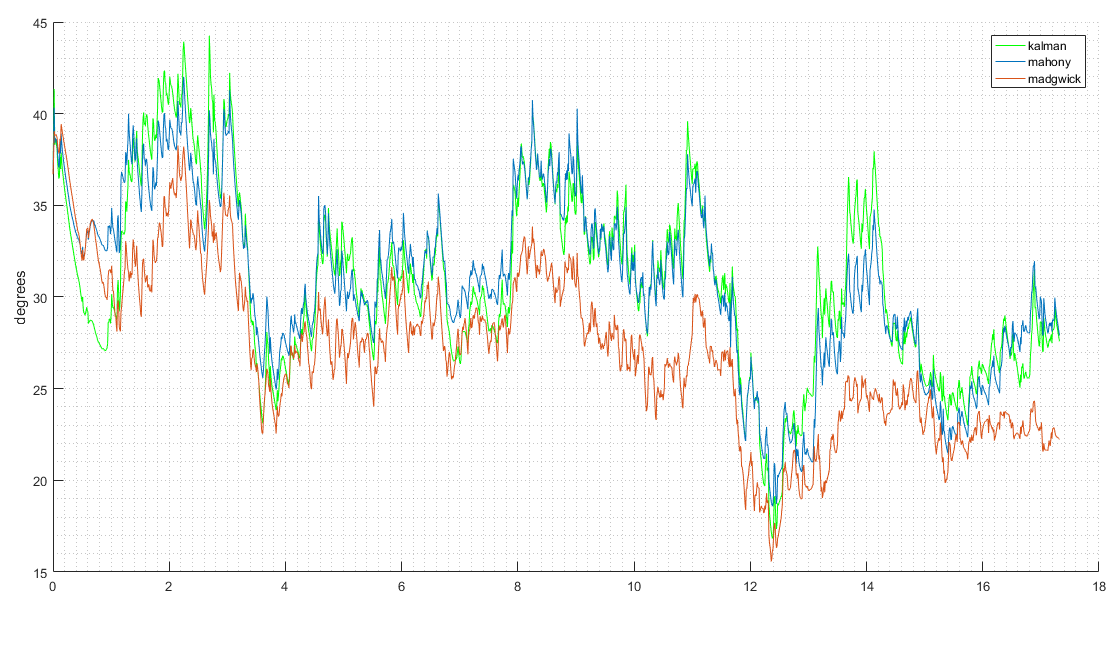}
\caption{Filters roll} 
\label{fig:roll}
\end{center}
\end{figure}

Different studies about the use of filters in IMU units data to improve the quality and reduce the noise in the data have been performed \cite{madgwick2011estimation} \cite{madgwick2010efficient} \cite{alam2014comparative}, showing that the Madgwick filter is the most appropriate in this kind of systems. In this paper, as complementary work we decided to implement tests of three of the most used filters, a Kalman filter, a Mahony filter and the aforementioned Madgwick filter. The representation of the pitch, yaw and roll obtained in the tests can be shown in Figure \ref{fig:yaw}, Figure \ref{fig:pitch} and Figure \ref{fig:roll}, respectively. The Kalman filter is represented in green color, the Mahony filter in blue and the Madgwick one in orange in the different plots. During the different tests, the Magdwick filter shows a better accuracy comparing the real position with the obtained in the Android application.

Finally, a step length estimation has been used to perform an exhaustive study of different methods. As initial method, we decided to use a simple way to calculate the step length in centimetres, $l$, which can be shown in the equation \ref{eq:stepLength}, where  $h$ represents the height in centimetres of the user and $k$ is a constant that is 0.415 for men and 0.413 for women \cite{bylemans2009mobile}. 

\begin{equation} 
\label{eq:stepLength}
 l = h \cdot k
\end{equation}

In future versions of the system, more efficient, precise and complex step length estimation methods will be implemented. Moreover, a comparative study of the accuracy of the different methods will be performed too and different test using the smartphone instead of the IMU will be performed.

The general system performance can be shown in Figure \ref{fig:workingMode}. The steps that a user of the system takes during its use are:

\begin{figure}[!ht]
\begin{center}
\includegraphics[width=1\linewidth]{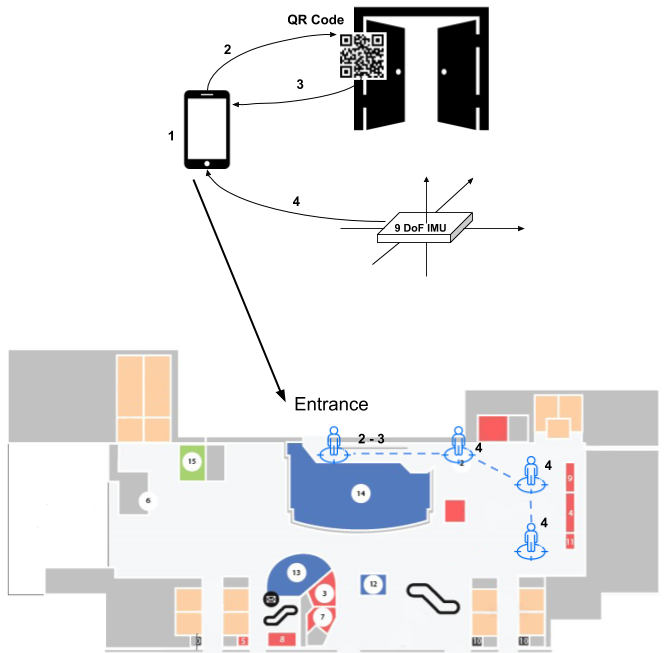}
\caption{General System Performance} \label{fig:workingMode}
\end{center}
\end{figure}

\begin{enumerate}
	\item The initial step of the system consists on putting the user's height the first time that he/she uses the application to calculate the step length. 
	\item The user scans the QR code situated in the airport entrance. The QR code contains some identification numbers that represents the building, the position and the floor. The possibility of put more QR codes around the building is open for some cases where the user forgot to do it at the entrance. This information is important to situate the user in the right place inside the building and floor. 
	\item Once  the QR code has been read, the user can see her/his initial position over the airport floor map. 
	\item At this moment, the IMU unit starts to collect data and send them to the user smartphone, which is  responsible for operating with it. The Madgwick algorithm is used to get the orientation in real time. With the quaternion obtained by algorithm and the step length, the user position for each step is shown in the smartphone over the airport map.
\end{enumerate}
	
\begin{figure*}[!ht]
\begin{center}
\includegraphics[width=1\linewidth]{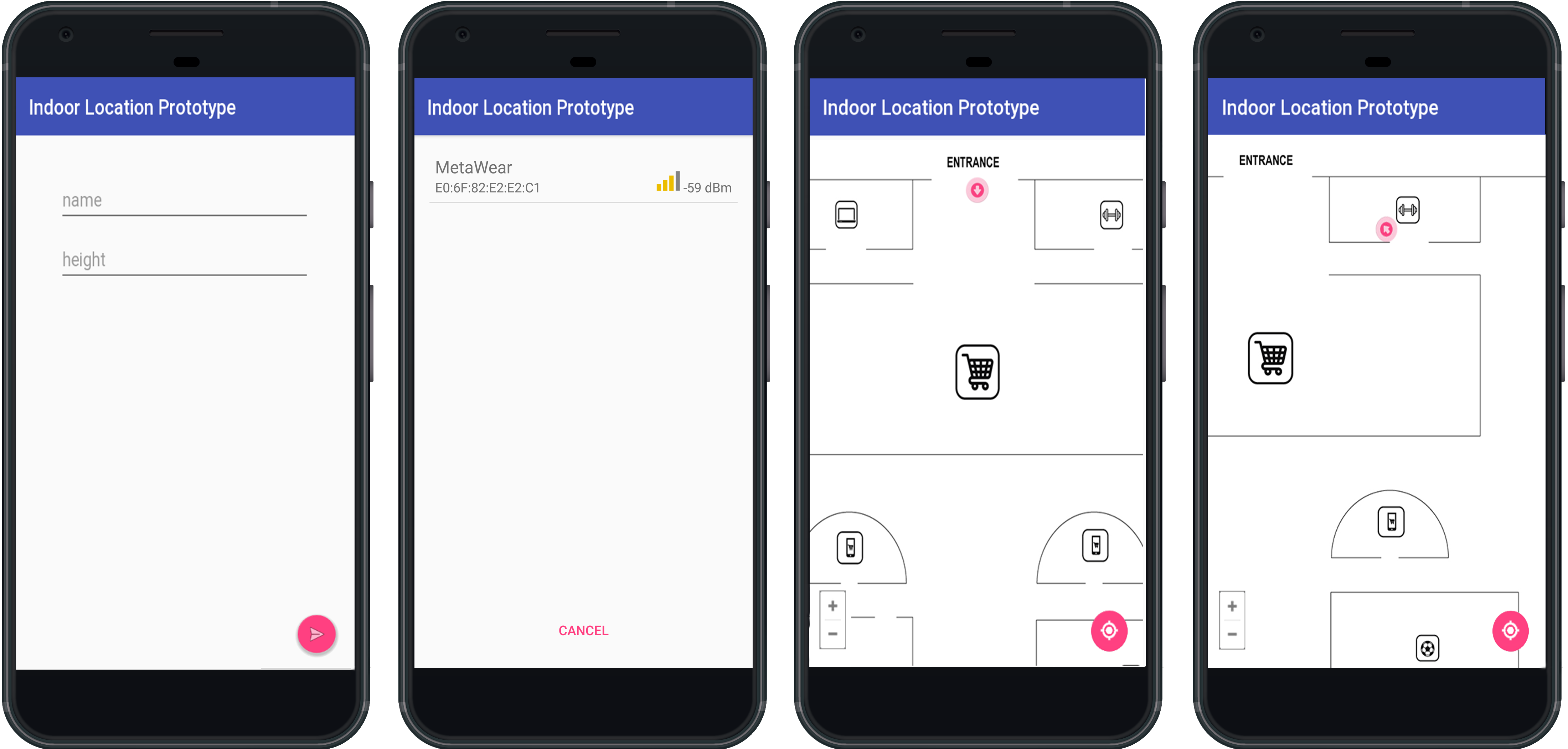}
\caption{Prototype Application} \label{fig:screenshots}
\end{center}
\end{figure*}

Some screenshots of the developed application using the aforementioned IoT devices can be shown in Figure \ref{fig:screenshots}.

\section{Security Scheme}
\label{sec:security}

The security in positioning and location systems is very important. A vulnerable application can involve a privacy problem, and in particular, a traceability one. In airports these kind of security problems could suppose that somebody could know what is your destination, track all your steps or simply know that somebody who lives in some direction is getting a flight to another place and for example steal these people house.

To protect the developed system, we decided to use a combination of two different algorithms, one to generate a shared key and another to encrypt all communications between smartphone and IMU. Moreover the generated keys are one time keys used as sessions keys, so every time these keys have to be renewed.

To generate the shared key used later to encrypt the communications an method based on elliptic curve cryptography has been selected. The use of elliptic curves in cryptography has been widely discussed during the last years and  the advantages that they have with respect to the traditional cryptography both in  key length and computational requirements are well known. The FourQ is a new elliptic curve developed by Microsoft Research \cite{liufourq}, which accomplishes the NIST requirements for the selection of new generation elliptic curves. These requirements are that  new curves must, at least, maintain the security level of the previous ones and  be highly efficient in  software and hardware implementations. This curve produces promising results, as shown in  different studies presented by Microsoft Research, and offers improvements in the computing times in the tests done in traditional computers. To know if the improvements shown in the Microsoft Research tests are possible too in portable devices, where the processor architecture is totally different, because  computers usually use an x32 or x64 architecture while the smartphones and portable devices usually use the arm architecture, we decided to port the implementation done by Microsoft Research to Java language to use it in Android devices. To do this, a java library was made and the FourQ computing time executing an Elliptic Curve Diffie-Hellman (ECDH) protocol was compared between FourQ, the NIST P-256 curve \cite{brown2001software} and the Curve25519 curve \cite{bernstein2006curve25519}. The results of this comparison can be seen in Table \ref{tab:fourQ}.

\begin{table}[!ht]
\caption{ECDH EXECUTION TIME COMPARISON}
\label{tab:fourQ}
\begin{center}
\begin{tabular}{ll}
\hline\noalign{\smallskip}
Curve 		& Time 									                      \\
\noalign{\smallskip}
\hline
\noalign{\smallskip}
Curve25519  						& 721 ms               \\
Curve NIST P-256 	&   1876 ms 				\\ 		
Curve FourQ 	& 417 ms		

\end{tabular}

\end{center}
\end{table}

The computing times shown in Table  \ref{tab:fourQ} show that the FourQ elliptic curve offers interesting improvements in portable devices too. In particular, this curve is 2 times faster than the new generation Curve25519 curve and around 4-5 times faster than the NIST P-256 curve. The use of this curve can  be an important advance in the IoT security due to the lower key length and  higher efficiency, facts that are specially important in this kind of devices with low computing  and  storage capacities.

\begin{figure}[!ht]
	\begin{center}
		\includegraphics[width=0.8\linewidth]{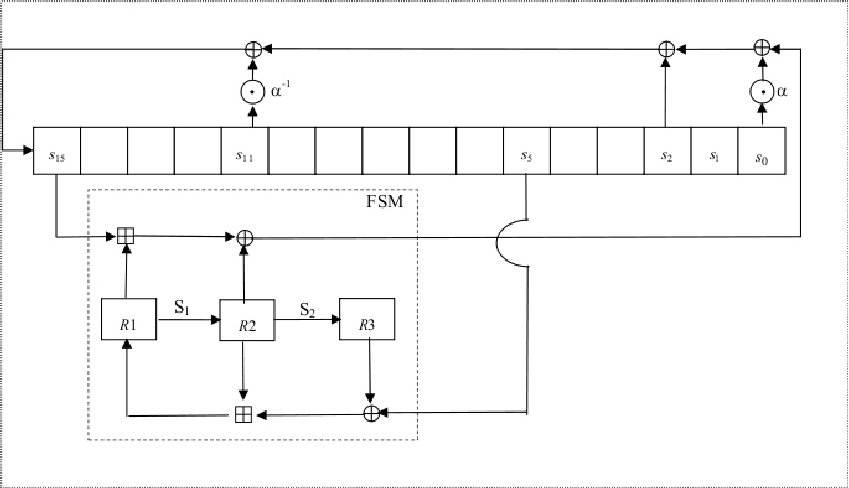}
		\caption{SNOW 3G scheme}
		\label{fig:snow}
	\end{center}
\end{figure}

Once the shared key has been established, the SNOW 3G algorithm is the responsible of encrypt all communications. The SNOW 3G is the stream cipher algorithm designated in 2006 as basis for the integrity protection and encryption of the UMTS technology. Thanks to the fact that the algorithm satisfies all the requirements imposed by the 3rd Generation Partnership Project (3GPP) with respect to time and memory resources, it was selected for the UMTS Encryption Algorithm 2 (UEA2) and UMTS Integrity Algorithm 2 (UIA2)  \cite{kitsos2008high} \cite{orhanou2010snow}.

The SNOW 3G algorithm derives from the SNOW 2 algorithm, and uses 128-bit keys and an initialization vector in order to generate in each iteration 32 bits of keystream.
On the one hand, the LFSR used in this algorithm has 16 stages denoted s0, s1, s2… s15 with 32 bits each one.
On the other hand, the used Finite State Machine (FSM) is based on three 32-bit records denoted R1, R2 and R3 and uses two Substitution-boxes called S1 and S2. The combination operation uses a XOR and an addition module $2^{32}$, as we can see in Figure \ref{fig:snow}. 

 SNOW 3G  has two execution modes: the initialization mode and the keystream mode. First, the initialization mode is executed without producing any keystream. Then, the keystream mode is executed. In particular, the number of iterations of such a mode depends on the number of 32-bit words that we want to generate. The performance of this algorithm can be observed in \cite{molina2014analysis}.

\section{Conclusions}
\label{sec:conclusions}

This work presents a novel indoor location and positioning system for airport environments that offers promising results. The presented solution uses two technologies to solve the problem of the GPS accuracy in this kind of scenarios. The use of the QR codes to provide an initial measurement point in combination with the IMU trough the use of users smartphones provides a low cost system in relation with the high precision that offers. An Android application to collect the IMU information, proceed with the different calculations and show the path over the airport indoor map has been developed. Different algorithms and security protocols have been implemented to ensure the security and no traceability of the users. This is a work in progress, so several lines are still open. The first of them is the use of the smartphone sensors to provide the information obtained from the IMU. The second other hand, the study of other sensor fusion algorithms that could fit  the developed system better coul be an important advance on the system. Finally, more security tests and controlled attacks to improve the system security are also necessary.

\section*{Acknowledgment}

Research supported by TESIS2015010102, TESIS-2015010106 and by the Spanish Ministry of Economy and Competitiveness, the European FEDER Fund, and the CajaCanarias Foundation, under Projects TEC2014-54110-R, RTC-2014-1648-8, MTM2015-69138-REDT and DIG02-INSITU.

\end{document}